  \documentclass[showpacs,showkeys,aps,prd,preprint,preprintnumbers,amsfonts,eqsecnum,nofootinbib,
floatfix,natbib]{revtex4-1}
\usepackage{color,graphicx,hyperref,dsfont,slashed,multirow}
\usepackage{placeins}
\usepackage{subfigure}
\usepackage{multirow,rotating}
\usepackage{color}
\usepackage{hyperref}

\hypersetup{bookmarks=true,unicode=true,pdftoolbar=true,pdfmenubar=true,
  pdffitwindow=false,pdfstartview={FitH},pdftitle={mBLRISS},
  pdfauthor={~M.~Frank,~K.~Huitu,~S.~Mondal}, pdfsubject={Subject},
  pdfcreator={Creator},pdfproducer={Producer}, pdfkeywords={Non-minimal 
  supersymmetry}{Beyond the Standard Model Physics}{non-universal Z' models},
  pdfnewwindow=true,colorlinks=true,
  linkcolor=blue,citecolor=magenta,filecolor=magenta,urlcolor=cyan}
\newcommand{\be}{\begin{equation}}
\newcommand{\ee}{\end{equation}}
\newcommand{\bea}{\begin{eqnarray}}
\newcommand{\eea}{\end{eqnarray}}
\def\bsp#1\esp{\begin{split}#1\end{split}}
\renewcommand{\figureautorefname}{Fig.}

\def\zpr{Z^{\prime}}
\def\mzpr{m_{Z^{\prime}}}

\def\mlsp{m_{\chi_1^0}}

\def\sectionautorefname~#1\null{Sec.~(#1)\null}
\def\subsectionautorefname~#1\null{sub--Sec.~(#1)\null}
\def\figureautorefname~#1\null{Fig.~#1\null}
\def\tableautorefname~#1\null{Table~#1\null}
\def\equationautorefname~#1\null{Eq.~#1\null}

\begin{document}
%\date{\today}
\preprint{CUMQ/HEP 200, HIP yyy}

{\title{Dark matter and Collider signals in supersymmetric $U(1)^\prime$ models with non-universal $Z^\prime$ couplings}

\author{Mariana Frank}
\email{mariana.frank@concordia.ca}
\affiliation{Department of Physics, Concordia University 7141 Sherbrooke St. West, Montreal, QC, CANADA H4B 1R6}

\author{Katri Huitu}
\email{katri.huitu@helsinki.fi}
\affiliation{Department of Physics, and Helsinki Institute of Physics, P. O. Box 64, FI-00014 University of Helsinki, Finland}

\author{Subhadeep Mondal}
\email{subhadeep.mondal@helsinki.fi}
\affiliation{Department of Physics, and Helsinki Institute of Physics, P. O. Box 64, FI-00014 University of Helsinki, Finland}

\date{\today}
\pacs{}
\keywords{Dark Matter,  $U(1)^\prime$, $Z^\prime$, flavor non-universality}%Use showkeys class option if keyword

\vspace{10pt}
\begin{abstract}
 
We analyse supersymmetric models augmented by an extra $U(1)$ gauge group. To avoid anomalies in these models 
without introducing exotics, we allow for family-dependent $U(1)^\prime$ charges, and choose a simple form for 
these, dependent on one $U(1)^\prime$ charge parameter only. With this choice, $Z^\prime$ decays into di-taus 
but not di-leptons, weakening considerably the constraints on its mass. 
In the supersymmetric sector, the effect is to lower the singlino mass, allowing it to be the dark matter candidate. 
We investigate the dark matter constraints and collider implications of such models, with mostly singlino, or mostly 
higgsinos, or a mixture of the two as lightest supersymmetric particles. In these scenarios, $Z^\prime$ decays 
significantly into chargino or neutralino pairs, and thus indirectly into final state leptons. We devise benchmarks 
which, with adequate cuts, can yield signals visible at the high-luminosity LHC.

\end{abstract}

%%%%%%%%%%%%%%%%

\maketitle
%%%%%%%%%%%%%%%%%
\section{Introduction}
\label{sec:intro}
%%%%%%%%%%%%%%%%%

Models with additional $U(1)$ gauge symmetries are a popular extension of the Standard Model (SM). Without supersymmetry, 
it was shown that they can provide a  model for dark matter \cite{Okada:2010wd,Okada:2016tci,Okada:2016gsh,Agrawal:2018vin},  
better agreement with  measurements of the anomalous magnetic moment of the muon \cite{Heeck:2011wj, Allanach:2015gkd},  
and explain leptogenesis \cite{Chen:2011sb}. In supersymmetry, they are motivated by the ability to generate  the $\mu$ 
parameter at the electroweak scale  
\cite{Fayet:1977yc,Komachenko:1989qn,Cvetic:1996mf,Suematsu:1994qm,Jain:1995cb,Nir:1995bu}.  If the extra $U(1)^\prime$ 
is a result of breaking of $E6$,  right-handed neutrinos emerging from the fundamental {\bf 27} fundamental representation 
can be incorporated into the model spectrum \cite{Keith:1996fv}.  An added benefit of supersymmetric models is 
that these explain the stability of the proton \cite{Carone:1996nd}, and provide fermion masses through the 
Froggatt-Nielsen mechanism \cite{Froggatt:1978nt}.

Extra $U(1)$ symmetries (which we shall refer to as $U(1)^\prime$ models) can arise as low-energy manifestations of  
grand unified theories \cite{Hewett:1988xc}, of string theories \cite{Cvetic:1995rj}, and from models with dynamical electroweak 
breaking \cite{Hill:2002ap}. In the framework of gauge mediation, they provide a mechanism for supersymmetry 
breaking \cite{Kaplan:1999iq}. A disadvantage of these models  is the requirement of cancellation of anomalies. 
Imposing that the theory be anomaly--free usually requires adding several exotics to the spectrum \cite{Erler:2000wu}, 
 introducing several new particles with respect to the minimal content, often spoiling the gauge coupling 
unification\footnote{Note however that coupling unification can be sometimes preserved, as in \cite{King:2007uj}, 
where, in the  $U(1)^\prime_N$ model, resulting from breaking supersymmetric $E6$, gauge unification is preserved, even in the 
presence of exotic remnants of $SU(5)$ representations.} 
a desirable  prediction of the minimal supersymmetric standard model (MSSM) with weak scale soft masses.

The goal of this work is to explore the consequences of an anomaly-free $U(1)^\prime$ model without 
additional exotic matter, without imposing that it be generated by the breaking of $SO(10)$ or $E6$. We also want 
to construct a model where we can relax the mass constraints on $\zpr$. Constructing anomaly-free $U(1)^\prime$ 
models without exotics is possible, but it  involves allowing flavor non-universality, 
that is, allowing fermions to have family-dependent $U(1)^\prime$ charges \cite{Demir:2005ti}. These charges must be chosen 
such that all anomaly coefficients cancel, including those from mixed anomalies involving $U(1)^\prime$ charges, and 
gauge-gravity anomalies. These particular theories have received more attention lately, given the LHCb measurements of lepton 
flavor non-universality in B-meson decays \cite{Aaij:2014ora,Aaij:2017vbb,Hiller:2003js}.

There are numerous possibilities for non-universal $U(1)^\prime$ charges. These are classified in \cite{Allanach:2018vjg} 
and various aspects of their phenomenological implications have been studied both within non-SUSY and SUSY 
frameworks \cite{Celis:2015ara,Allanach:2019mfl,Alvarado:2019gyh,Mantilla:2016lui,Tang:2017gkz,Kamenik:2017tnu,Coleppa:2018fau}. 
In this work, we revisit the supersymmetric $U(1)^\prime$ models with non-universal charges, opting for a simple family dependent choice. Our 
aim is to study the phenomenology of the $Z^\prime$ boson, which in these scenarios can be light\footnote{In models with universal 
$U(1)^\prime$ charges, $\zpr$ masses are restricted rather stringently by the ATLAS  \cite{Aaboud:2017yvp} and 
CMS \cite{CMS:2016abv} collaborations, and expected to be around 4 - 4.5 TeV. These models can be rendered leptophobic 
by using kinetic mixing between the two $U(1)$ gauge groups, as in {\it e.g.} \cite {Araz:2017wbp}.}. In addition to 
consequences observable at colliders, $Z^\prime$ mass plays a role in fine-tuning, rendering scenarios with low $\zpr$ mass  
interesting theoretically. We explore how restrictive is the $Z^\prime$ mass, and the signatures of such a boson at the colliders.

Related to these, we  also investigate the phenomenology of dark matter in these models \cite{Araz:2017qcs,Frank:2014bma,
Hicyilmaz:2016kty,Darme:2018hqg}, with emphasis on effects of a lighter 
$\zpr$, and on the possibility of having the singlino (the fermion partner of the singlet Higgs boson required to break 
$U(1)^\prime$ symmetry), as a non-standard dominant component of dark matter. 

As an artefact of allowing flavor non-universality, the $\zpr$ phenomenology at the LHC can be quite distinctive. The $\zpr$ can now 
decay into certain favored final states dominantly, while some of the more commonly observable decay modes are absent altogether. Here,  
 one possible solution to the various anomaly-cancellation equations leads to a scenario where the $\zpr$ is 
forbidden to decay into electron or muon pairs. Instead, its single most prominent decay mode is $\tau\bar\tau$. Naturally, in this scenario, the existing 
constraints on $\zpr$ mass can be quite relaxed. On the other hand, within a SUSY framework, there can be additional decay modes of the $\zpr$ 
which may lead to hitherto unexplored signal regions. We have explored two such signal regions and present our results in the context of high 
luminosity run of the LHC at a centre-of-mass energy of 14 TeV. 

The paper is organised in the following way. In section~\ref{sec:model} we describe briefly the theoretical framework of our study. In 
section~\ref{sec:lhc_const} we discuss the impact of various LHC search results on the parameter space of our model. Based on that study,
we proceed to select some representative benchmark points. In section~\ref{sec:coll} we discuss the strategy to explore these classes of 
benchmark points at the 14 TeV LHC. We discuss  possible signal regions, SM background contributions and kinematic cuts that can be 
used to suppress these background contributions and make the signal observable. We discuss our results through detailed cut-flow tables 
and finally conclude our observations in section~\ref{sec:concl}.

%%%%%%%%%%%%%%%%%
\section{The $U(1)^\prime$ model with non-universal charges}
\label{sec:model}

Supersymmetric $U(1)^\prime$ models are based on the gauge group $SU(3)_c\otimes SU(2)_L \otimes U(1)_Y\otimes U(1) ^{\prime}$, with
gauge couplings $g_s$, $g_2$, $g_Y$ and $g^{\prime}$\footnote{The $SU(2)_L\times U(1)_Y\times U(1)^{\prime}$ covariant derivative 
is given by $D_{\mu}=\partial_{\mu} + ig_2T^aW_{\mu}^a + ig_Y YV_{\mu} + ig^{\prime}Y^{\prime}V_{\mu}^{\prime}$}. 
The particle spectrum of
the models is that of the MSSM augmented by a gauge singlet $S$, charged
under $U(1)^\prime$ only. The particle content, allowing for non-universal charges under the $U(1)^\prime$ group, is given in 
Table \ref{tab:particles}.
\begin{table}[h]
\begin{center}
\begin{tabular}{|c|c|c|c|c|} \hline
      & SU(3)$_c$ & SU(2)$_L$  & U(1)$_Y$ & U(1)$^{\prime}$   \\ \hline
$Q_i$   & ${\bf 3}$       &  ${\bf 2}$       &  $1/6$   & $Q_{Q_i}$ \\ \hline
$U^c_i$   & $\bar {\bf 3}$  &  ${\bf 1}$       &  $-2/3$  & $Q_{U^c_i}$ \\ \hline
$D^c_i$   & $\bar {\bf 3}$  &  ${\bf 1}$       &  $1/3$   & $Q_{D^c_i}$ \\ \hline
$L_i$   & ${\bf 1}$       &  ${\bf 2}$       &  $-1/2$  & $Q_{L_i}$ \\ \hline
$E^c_i$   & ${\bf 1}$       &  ${\bf 1}$       &  $1$     & $Q_{E^c_i}$ \\ \hline
$H_u$ & ${\bf 1}$       &  ${\bf 2}$       &  $1/2$   & $Q_{H_u}$ \\ \hline
$H_d$ & ${\bf 1}$       &  ${\bf 2}$       &  $-1/2$  & $Q_{H_d}$ \\ \hline
$S$   & ${\bf 1}$       &  ${\bf 1}$       &  $0  $   & $Q_{S}$ \\ \hline
\end{tabular}
\end{center}
\caption{The particle content of the $U(1)^\prime$ model, and assignments under the different groups, allowing for different 
charges under the $U(1)^\prime$ group. The index $i$ runs over three families.}
\label{tab:particles}
\end{table}

The breaking of the $U(1)^\prime$ gauge symmetry down to electromagnetism is achieved through the
neutral components of the scalar Higgs fields acquiring  VEVs,
$\langle H^0_u\rangle = v_u/\sqrt{2}$, $\langle H^0_d\rangle = v_d/\sqrt{2}$ and
$\langle S\rangle = v_S/\sqrt{2}$.

The superpotential takes the form:
\be \widehat{W}= \lambda \widehat{S} \widehat{H}_d \widehat{H}_u + h_u^{ij} \widehat{U}^c_{j} \widehat{Q}_i \widehat{H}_u 
+ h_d^{ij} \widehat{D}^c_j \widehat{Q}_i
\widehat{H}_d+ h_e^{ij} \widehat{E}^c_j \widehat{L}_i
\widehat{H}_d\,. 
\label{eq:superpotential}
\ee
Here the first term of the superpotential is responsible for generating an 
effective $\mu$ parameter $\lambda \langle S \rangle$, providing a dynamical solution to the $\mu$
problem when $\langle S \rangle \sim {\cal{O}}({\rm TeV})$. The
rest of the operators in (\ref{eq:superpotential}) are the usual Yukawa terms 
interactions of leptons and quarks. 

The most general holomorphic Lagrangian responsible for soft supersymmetry breaking is

\bea 
-{\mathcal{L}}_{soft}&=&\left(\sum_i M_i \lambda_i\lambda_i- A_{\lambda}\lambda SH_dH_u
 -A_u^{ij}h_u^{ij} U^c_jQ_iH_u-A_{d}^{ij}h_d^{ij} D^c_jQ_iH_d-
 A_e^{ij}h_e^{ij} E^c_jL_iH_d+h.c.\right)  \nonumber \\
 &+& m_{H_u}^2|H_u|^2+ m_{H_d}^2|H_d|^2+m_S^2|S|^2+ 
  m_{Q_{ij}}^2 \widetilde Q_i\widetilde Q_j^*
    +m_{U_{ij}}^2 \widetilde U^c_i\widetilde U^{c*}_j+m_{D_{ij}}^2 \widetilde D^c_i\widetilde D^{c*}_j
    +m_{L_{ij}}^2 \widetilde L_i\widetilde L_j^*\nonumber \\
    &+& m_{E_{ij}}^2 \widetilde E^c_i\widetilde E^{c*}_j+h.c. \, ,
    \label{eq:soft}
\eea
where the SUSY-breaking sfermion mass-squared $m_{Q,\dots,E^c}^2$ and the trilinear couplings $A_{u,\dots,e}$ are $3\times 3$ 
matrices in flavor space, and are assumed here to be diagonal, while gaugino masses $M_i$ and trilinear couplings $A_{S, \dots, e}$ 
are taken to be real. 

Family dependent $U(1)^{\prime}$ charge assignment forbids some of the Yukawa couplings in the superpotential, resulting in 
massless fermions. One therefore must introduce non-holomorphic SUSY breaking Lagrangian, induced by the couplings of fermions 
to the 'wrong' Higgs doublet 
\be
  -{\mathcal{L}}_{c}=
C_E^{ij}H_{u}^* \widetilde L^i \widetilde E_R^{cj}+
 C_U^{ij} H_{d}^* \widetilde Q^i\widetilde U_R^{cj} +
 C_D^{ij} H_{u}^*\widetilde Q^i\widetilde D_R^{cj} + c.c.\, ,
\label{eq:nonholonomic}
\ee  
which is essential for giving mass to fermions. The fermion masses are generated at one loop level through sfermion-gaugino loops
\cite{Demir:2005ti}.

For the theory to be anomaly-free, the $U(1)^\prime$ charges must satisfy conditions requiring  vanishing of 
$U(1)^{\prime}-SU(3)-SU(3)$, $U(1)^{\prime}-SU(2)-SU(2)$, $U(1)^{\prime}-U(1)_Y-U(1)_Y$, $U(1)^{\prime}$-graviton-graviton,
$U(1)^{\prime}-U(1)^{\prime}-U(1)_Y$ and $U(1)^{\prime}-U(1)^{\prime}-U(1)^{\prime}$ anomalies, that is, the charges must satisfy, respectively
\bea 
0&=&\sum_i(2Q_{Q_i}+Q_{U^c_i}+Q_{D_i}) \\
0&=&\sum_i(3Q_{Q_i}+Q_{L_i})+Q_{H_d}+Q_{H_u} \\
0&=&\sum_i(\frac{1}{6}Q_{Q_i}+\frac{1}{3}Q_{D^c_i}+
 \frac{4}{3}Q_{U^c_i}+\frac{1}{2}Q_{L_i}+Q_{E^c_i})+
 \frac{1}{2}(Q_{H_d}+Q_{H_u}) \\
0&=&\sum_i(6Q_{Q_i}+3Q_{U^c_i}+3Q_{D^c_i}+2Q_{L_i}+Q_{E^c_i})+2Q_{H_D}
 +2Q_{H_u}+Q_s \\ 
0&=&\sum_i(Q_{Q_i}^2+Q_{D^c_i}^2-2Q_{U^c_i}^2-Q_{L_i}^2+Q_{E^c_i}^2)-
 Q_{H_d}^2+Q_{H_u}^2 \\ 
0&=&\sum_i(6Q_{Q_i}^3+3Q_{D^c_i}^3+3Q_{U^c_i}^3+2Q_{L_i}^3+Q_{E_i}^3)+
 2Q_{H_d}^3+2Q_{H_u}^3+Q_S^3.
\label{eq:anomaly}
\eea
A possible solution to the above, satisfying the anomaly cancellation requirement is
\bea
&&Q_{E_{1,2}}=Q_{L_{1,2}}=Q_{L_3}=0 \nonumber \\
&&Q_{Q_i}=\frac{Q_{E_3}}{9}\,  ; \, Q_{D_i}=-\frac{Q_{E_3}}{9}\, ; \, Q_{U_i}=-\frac{Q_{E_3}}{9}\, ; \nonumber \\
&&Q_{H_u}=0\, ; \, Q_{H_d}=-Q_{E_3}\, ; \, Q_{S}=Q_{E_3}\, ,
\label{eq:Qchoices}
\eea
which is by no means general, but allows us to express all $U(1)^\prime$ charges in terms of a single one, $Q_{E_3}$.

%%%%%%%%%%%%%%%%%%%%%%%%%%%%%%
%\subsection{Scalar sector}
%\label{sec:scl}
%%%%%%%%%%%%%%%%%%%%%%%%%%%%%%
%In this framework, the neutral scalar mass matrix is of the following form
%\bea 
%\left(\begin{array}{ccc}
%	\frac{1}{4}(g_1^2+g_2^2+4{g^{\prime}}^2Q_{H_d}^2)v_d^2 + \frac{T_{\lambda}v_Sv_u}{\sqrt{2}v_d} 
%	& -\frac{1}{4}(g_1^2+g_2^2-4\lambda^2)v_dv_u - \frac{T_{\lambda}v_S}{\sqrt{2}} 
%	& (\lambda^2+{g^{\prime}}^2Q_{H_d}Q_S)v_dv_S - \frac{T_{\lambda}v_u}{\sqrt{2}}\\
%        -\frac{1}{4}(g_1^2+g_2^2-4\lambda^2)v_dv_u - \frac{T_{\lambda}v_S}{\sqrt{2}} 
%        & \frac{1}{4}(g_1^2+g_2^2)v^2_u + \frac{T_{\lambda}v_dv_S}{\sqrt{2}v_u} 
%        & \lambda^2v_Sv_u -\frac{T_{\lambda}v_d}{\sqrt{2}} \\
%        (\lambda^2+{g^{\prime}}^2Q_{H_d}Q_S)v_dv_S - \frac{T_{\lambda}v_u}{\sqrt{2}} 
%        & \lambda^2v_Sv_u -\frac{T_{\lambda}v_d}{\sqrt{2}} 
%        & {g^{\prime}}^2Q^2_Sv^2_S + \frac{T_{\lambda}v_dv_u}{\sqrt{2}v_S}  \nonumber
%      \end{array}\right) \\
%\eea

%%%%%%%%%%%%%%%%%%%%%%%%%%%%%%
\subsection{Neutralino sector}
\label{sec:neut}
%%%%%%%%%%%%%%%%%%%%%%%%%%%%%%
In this framework, the neutralino mass matrix in the basis $(\lambda_U,\lambda_{\widetilde B},\widetilde W,\widetilde{H}_d^0,\widetilde{H}_u^0,\widetilde{S})$ is 
of the following form
\bea 
\left(\begin{array}{cccccc}
	M_4 & 0 & 0 & g^{\prime}Q_{H_d}v_d& g^{\prime}Q_{H_u}v_u& g^{\prime}Q_{S}v_S\\
        0 & M_1 & 0 & -\frac{1}{2}g_1 v_d& \frac{1}{2}g_1 v_u& 0\\
        0 & 0 & M_2 & \frac{1}{2}g_2 v_d& -\frac{1}{2}g_2 v_u& 0\\
        g^{\prime}Q_{H_d}v_d & -\frac{1}{2}g_1 v_d & \frac{1}{2}g_2 v_d & 0& -\frac{1}{\sqrt{2}}v_S\lambda& -\frac{1}{\sqrt{2}}v_u\lambda\\
        g^{\prime}Q_{H_u}v_u & \frac{1}{2}g_1 v_u & -\frac{1}{2}g_2 v_u & -\frac{1}{\sqrt{2}}v_S\lambda& 0& -\frac{1}{\sqrt{2}}v_d\lambda\\
        g^{\prime}Q_S v_S & 0 & 0 & -\frac{1}{\sqrt{2}}v_u\lambda& -\frac{1}{\sqrt{2}}v_d\lambda & 0 \nonumber
     \end{array}\right) \\ 
\eea
%%%%%%%%%%%%%%%%%%%%%%%%%%%%%%
It is evident from the neutralino mass matrix that the LSP can be singlino dominated only if $g^{\prime}Q_S v_S$ is small enough 
and $M_4$ is heavy enough to be decoupled from the singlino mass. When all other soft masses are decoupled and there is almost zero 
mixing, the singlino mass is simply driven the the parameters $g^{\prime}$, $Q_S$ and $v_S$.  
These parameters also drive the $\zpr$ mass and as a result, if one looks for a light $\zpr$, a light singlino is always obtained. Depending on 
the choice of $\lambda$, the higgsinos can be light as well. In our present study, we have kept $M_1$, $M_2$ and $M_4$ heavy enough such that 
they decouple from rest of the spectra. 

%%%%%%%%%%%%%%%%%%%%%%%%%%%%%%
\section{Constraints}
\label{sec:lhc_const}
%%%%%%%%%%%%%%%%%%%%%%%%%%%%%%
LHC collaborations have explored various signal regions for any possible hint of a $\zpr$. The most stringent constraint 
is derived from high-mass dilepton resonance searches which exclude $\zpr$ mass ($\mzpr$) up to 4.5 TeV from data accumulated at 
$\sqrt{s}=13$ TeV with $139~{\rm fb}^{-1}$ luminosity \cite{Aad:2019fac}. Search for heavy particles decaying into a 
top-quark pair results in an exclusion limit on $\mzpr$ ranging from 3.1 TeV to 3.6 TeV at $\sqrt{s}=13$ TeV with 
$36.1~{\rm fb}^{-1}$ luminosity \cite{Aaboud:2019roo}. Dijet resoance search limit on $\mzpr$ is slightly weaker, 
$\mzpr > 2.7$ TeV at $\sqrt{s}=13$ TeV with $36~{\rm fb}^{-1}$ luminosity \cite{Sirunyan:2018xlo}. Thus it is evident 
that the most stringent constraint on $\mzpr$ is derived from its leptonic decay modes. Consequently, models with a  
leptophobic $\zpr$ \cite{Babu:1996vt,Suematsu:1998wm,Chiang:2014yva,Araz:2017wbp} are much less constrained in comparison. $\zpr\to\tau\tau$ decay deserves a special mention 
in this regard since the $\tau$ can decay both leptonically and hadronically. A combined search of both leptonically and 
hadronically decaying $\tau$-pairs exclude $\mzpr$ up to 2.42 TeV at $\sqrt{s}=13$ TeV with $36~{\rm fb}^{-1}$ luminosity 
\cite{Aaboud:2017sjh}. 

These existing exclusion limits are expected to vary depending on the assignments of $U(1)^{\prime}$ charges ($Q$) since these affect 
the production cross-section of $Z^{\prime}$. 
In the present scenario, $\zpr$ is forbidden to decay into light lepton pairs at the tree level. The $\zpr$ therefore, mostly decays via a pair 
of $\tau$-leptons. This, along with the decay into a neutralino/chargino pair accounts for most of the $\zpr$ width. 
Thus apart from the direct search limit on $\mzpr$, an indirect limit can also be derived from chargino/neutralino search results. 
This {\it new} decay mode of the $\zpr$ can contribute to the multilepton signal rate at the LHC. A $\zpr$ search in such 
signal regions has not been performed. 

Indirect constraints can be derived on $\mzpr$ from dark matter requirements. In this work we will focus on singlino and higgsino 
LSP scenarios. A pure singlino LSP can only annihilate efficiently around the Higgs and $\zpr$ resonances. However, the Higgs resonance 
region can be safely ruled out from LHC constraints on $\mzpr$. The $\zpr$ resonance region depends on the choice of model parameters. 
It is therefore worth checking if one can obtain a sub-TeV singlino DM in the present framework and still be consistent with the 
exclusion limits on $\mzpr$. Relic density requirement forces a pure higgsino DM to lie above 1 TeV. LSP higgsino masses below that
yield relic underabundance due to too much co-annihilation \cite{Kowalska:2018toh}. Direct search limits on the higgsino mass under
such circumstances are weak, around 200 GeV at $\sqrt{s}=13$ TeV with $139~{\rm fb}^{-1}$ luminosity \cite{Aaboud:2017leg}. 

%%%%%%%%%%%%%%%%%%%%%%%%%%%%%%
In order to understand the relevant parameter space, we have carried out detailed scans of the parameter space. 
The model was implemented in SARAH-4.14.0 \cite{Staub:2008uz,Staub:2009bi,Staub:2010jh,Staub:2013tta,Staub:2015kfa} which does the 
analytical calculation and writes the required files for implementing the 
model in numerical packages SPheno-4.0.2 \cite{Porod:2003um,Porod:2011nf,Porod:2002wz} and MicrOMEGAs-4.3.5 \cite{Belanger:2013oya}. 
SPheno calculates the masses, mixing matrices and the decay branching ratios of all the particles. MicrOMEGAs is used for the DM computations. 
We intend to explore both the singlino and higgsino LSP scenarios and hence we divide our scans into small $\lambda$ and large $\lambda$ cases. 
 
The $\lambda$ parameter multiplied by the singlet VEV generates the effective $\mu$-term in this mode, and  therefore drives the higgsino masses. 
As seen from the neutralino mass matrix, the $\lambda$ parameter also impacts the singlino-higgsino mixing. Therefore, when the $\lambda$ parameter 
is larger, one obtains a large parameter space where the LSP is a pure singlino and the higgsinos are heavier than the $\zpr$. In this  case, the 
$\zpr$  decays dominantly into a tau-pair and hence this parameter space is more likely to be excluded by the di-tau search channel. On the other hand, 
when the $\lambda$ parameter is smaller, the LSP can be a singlino-higgsino admixture or even a pure higgsino one. Apart from the LSP, there can also 
be additional chargino-neutralino states lying below the $\zpr$ and,  in the presence of these decay modes, its decay branching ratio into the tau-pair 
is reduced. For this case, the multilepton final state is quite relevant. The benchmark points chosen  reflect these facts. We fix all the $U(1)^{\prime}$ 
charges by $Q_{E_3}$ and note that these charges always appear together with the coupling $g^{\prime}$. We therefore, consider $g^{\prime}Q_{E_3}$ as one 
single parameter to vary. Below are the parameters ranges we consider: 
\begin{eqnarray}
g^{\prime}Q_{E_3}\equiv [0.01:0.9]; ~~{\rm tan}\beta\equiv [5.0:15.0]; \nonumber \\
v_S\equiv [1.0:15.0]~{\rm TeV}, ~~A_{\lambda}\equiv [1.0:15.0]~{\rm TeV}.
\end{eqnarray}
We have randomly generated points within these parameter ranges. Overall, we have generated about 100,000 points for each scan. 
Points are then passed through the constraints like 125 GeV Higgs mass, its coupling strengths with standard model particles and flavor constraints. 
The surviving points are shown in the subsequent figures.
%%%%%%%%%%%%%%%%%%%%%%%%%%%%%%%
\subsubsection{Large $\lambda$}
%%%%%%%%%%%%%%%%%%%%%%%%%%%%%%%
Throughout this scan we keep $\lambda = 0.6$, $M_1 = M_2 = M_4 = 4$ TeV. All the slepton and squark masses are kept at or above 3 TeV.
The exclusion limits, as obtained, are shown in Fig.~\ref{fig:excl_mzp_lam06}. The color gradient represents 
either the variation of the LSP neutralino mass or $g^{\prime}Q_{E_3}$,  as indicated in the figure. 
%%%%%%%%%%%%%%%%%%%%%%%%%%%%
\begin{figure}
\begin{center}
\includegraphics[width=6.2cm,height=6.2cm]{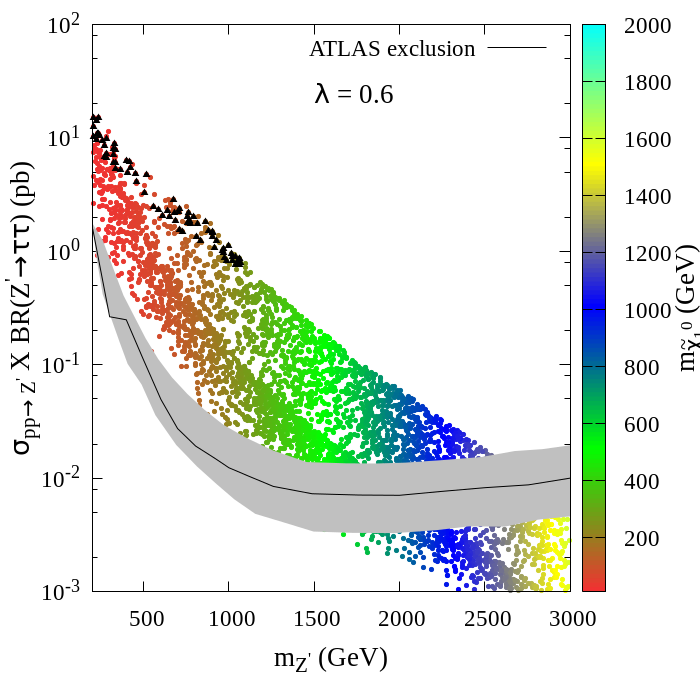}
\includegraphics[width=6.2cm,height=6.2cm]{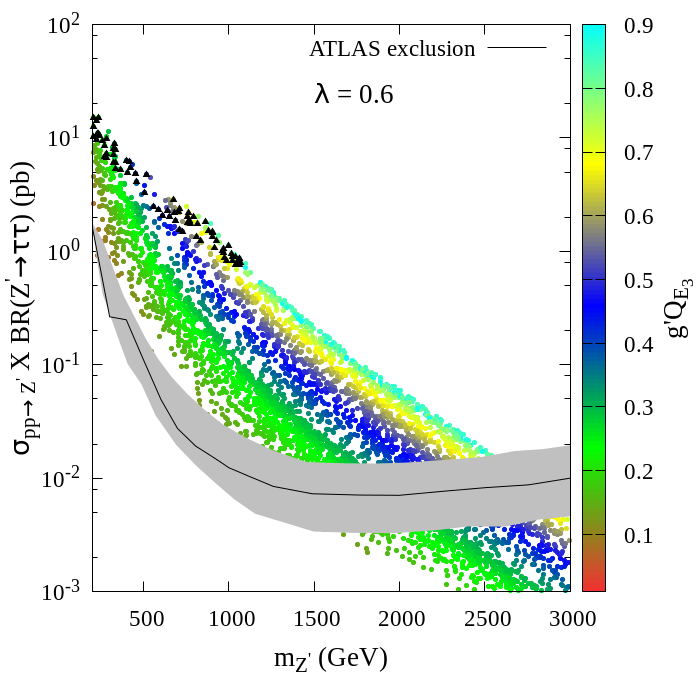}
\end{center}
\caption{Impact of the existing exclusion limit on $\mzpr$ from $\zpr\to\tau\tau$ search channel with $\lambda=0.6$.
The color coding represents the variation of the LSP neutralino mass or $g^{\prime}Q_{E_3}$ as indicated in the respective plots. The black points 
indicate those points that are ruled out from direct chargino-neutralino searches. The grey shaded region represents $95\%$ exclusion region 
around the observed limit.}
\label{fig:excl_mzp_lam06}
\end{figure}
%%%%%%%%%%%%%%%%%%%%%%%%%%%%
The exclusion limit obtained from $\zpr\to\tau\tau$ search is shown by the black line while the grey shaded area represents the $95\%$ 
confidence level region around the exclusion line \cite{Aaboud:2017sjh}. The black points represents those excluded from direct 
neutralino-chargino searches \cite{Aad:2014vma,Aaboud:2017leg,Aaboud:2018sua}. These constraints do not appear to affect the available 
parameter region significantly. This is because  the $\lambda$ parameter is relatively large which ensures that the 
higgsino mass parameter is quite large compared to the singlino in most of the cases. The bino and wino parameters  being also large 
throughout, both the chargino states and other neutralino states in the spectrum are quite heavy and   the singlino is the LSP 
state, which can still be significantly light. Thus the NLSP pair or the LSP-NLSP associated production cross-sections are very small. 
On the other hand, the LSPs can be produced copiously, but they are completely invisible. As expected, the exclusion limit on $\mzpr$ become weaker as 
$g^{\prime}Q_{E_3}$ is decreased since the production cross-section drops with it. As evident, with $g^{\prime}Q_{E_3}\sim 0.2$, 
the exclusion limit can be much weaker, $\mzpr\gtrsim 1500$ GeV. 

Now let us look at the DM properties. The distribution of the relic density as a function of the LSP neutralino mass is shown in 
Fig.~\ref{fig:relic_mch_lam06}. The color coding in the plots from left to right indicate the variation of $\mzpr$, the abundance of singlino component 
in the LSP and the relic density respectively. The horizontal shaded band represents $2\sigma$ allowed region around 
the correct relic abundance, $0.119\pm 0.0054$ \cite{Hinshaw:2012aka}. The XENON limit \cite{Aprile:2018dbl} on the direct detection cross-section 
($\sigma_{SI}$) is shown by the 
black curve. The two distinct resonance regions shown in the figure are due to the two CP-even Higgs masses corresponding to the MSSM Higgs doublets. 
For small $g^{\prime}Q_{E_3}$ the LSP is dominantly singlino resulting in very small $\sigma_{SI}$ which increases as the LSP becomes a singlino-higgsino 
admixture.  
%%%%%%%%%%%%%%%%%%%%%%%%%%%%
\begin{figure}
%\begin{center}
\subfigure[]{
\includegraphics[width=5.2cm,height=5.5cm]{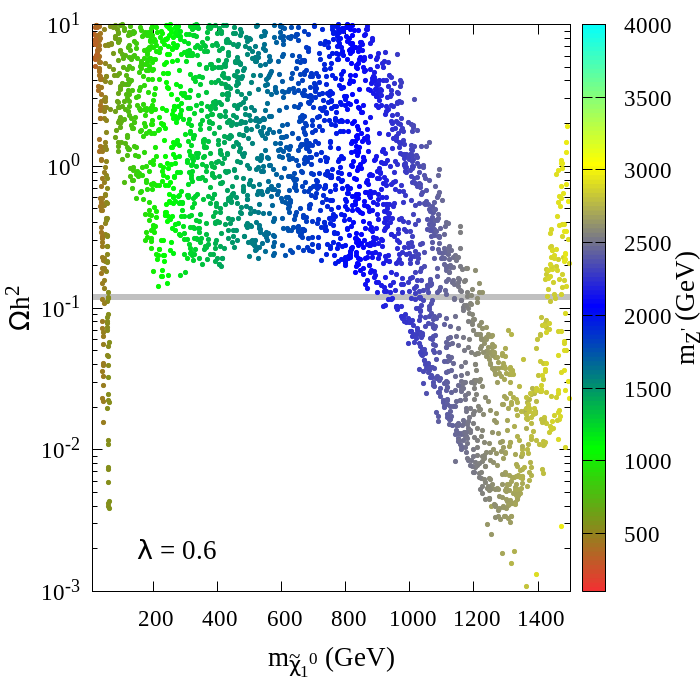}}
\subfigure[]{
\includegraphics[width=5.2cm,height=5.5cm]{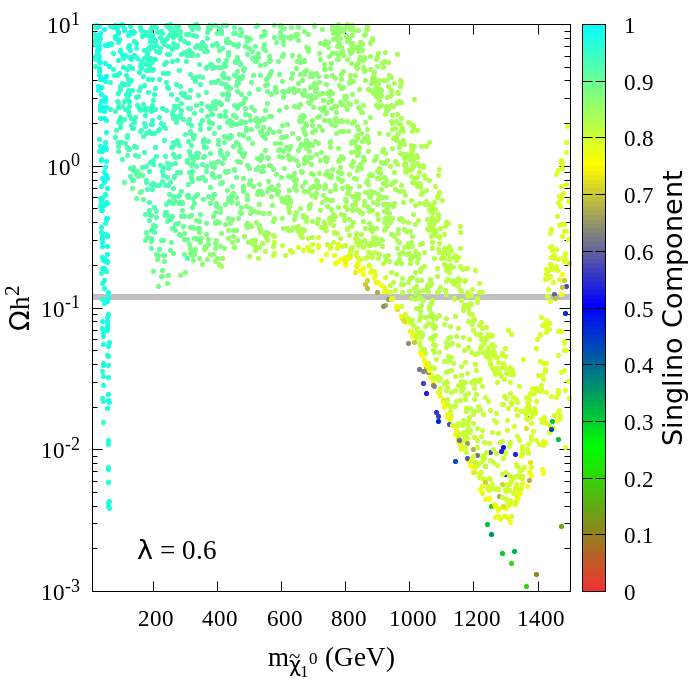}}
\subfigure[]{
\includegraphics[width=5.2cm,height=5.5cm]{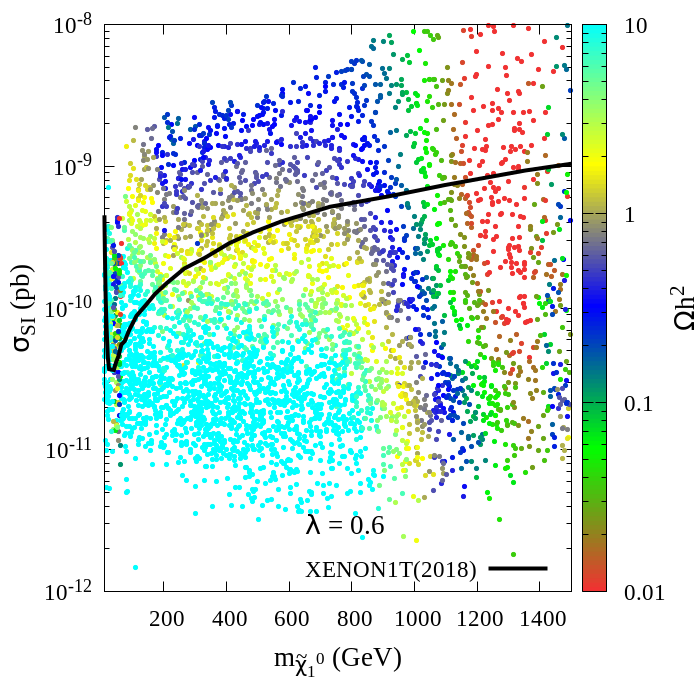}}
\caption{Distribution of the relic density as a function of the LSP mass with the color gradient representing variation of  
$\mzpr$ (a) and the abundance of the singlino component in the LSP neutralino (b). Plot (c) shows the distribution of 
direct detection cross-section as a function of LSP mass with the color gradient representing relic density. 
All the distributions are for $\lambda=0.6$.} 
\label{fig:relic_mch_lam06}
\end{figure}
%%%%%%%%%%%%%%%%%%%%%%%%%%%%
The admixture of singlino and higgsino produces more underabundance of relic density below 1.4 TeV, yielding  a wider range of parameter 
space satisfying the relic density requirement. 
Note that, the LHC limit on $\mzpr\gtrsim 2$ TeV rules out a significant portion of the DM allowed parameter region as indicated by blue points 
in Fig.~\ref{fig:relic_mch_lam06}(a). 
%%%%%%%%%%%%%%%%%%%%%%%%%%%%%%%
\subsubsection{Small $\lambda$}
%%%%%%%%%%%%%%%%%%%%%%%%%%%%%%%
Throughout this scan we keep $\lambda = 0.1$, $M_1 = M_2 = M_4 = 4$ TeV. All the slepton and squark masses are kept at or above 3 TeV.
The exclusion limits, as obtained, are shown in Fig.~\ref{fig:excl_mzp_lam01}. The color gradient 
represents either the variation of the LSP neutralino mass or $g^{\prime}Q_{E_3}$ as indicated in the figure. The exclusion limit is taken 
from the most recent results published by ATLAS collaboration \cite{Aaboud:2017sjh}. 
%%%%%%%%%%%%%%%%%%%%%%%%%%%%
\begin{figure}
\begin{center}
\includegraphics[width=6.2cm,height=6.2cm]{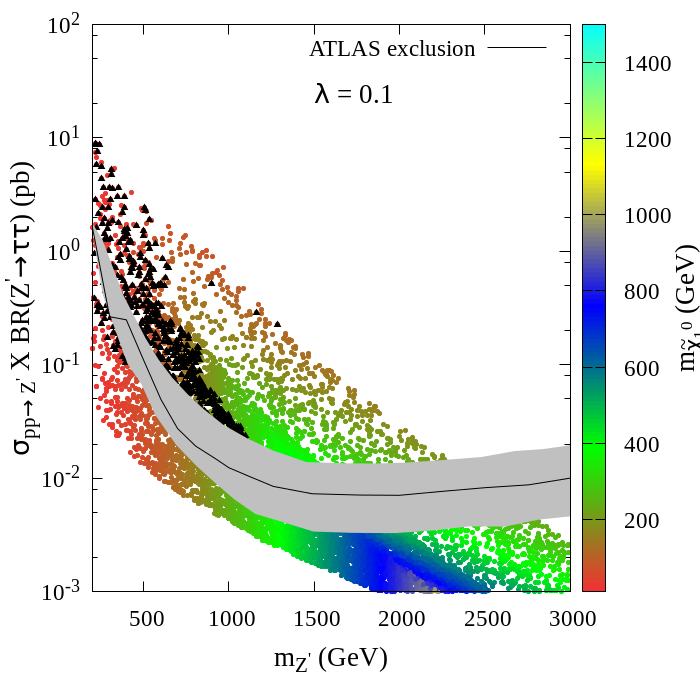}
\includegraphics[width=6.2cm,height=6.2cm]{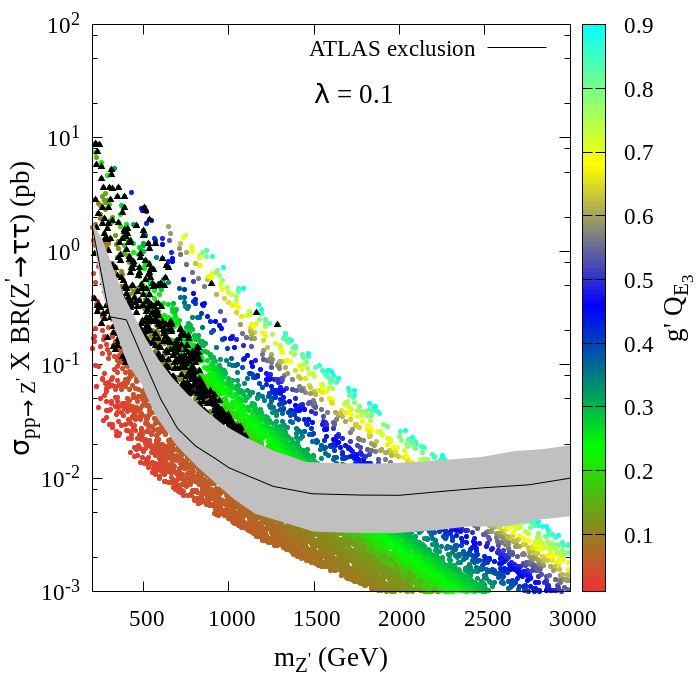}
\end{center}
\caption{Impact of the existing exclusion limit on $\mzpr$ from $\zpr\to\tau\tau$ search channel with $\lambda=0.1$. 
The color coding represents the variation of the LSP neutralino mass or $g^{\prime}Q_{E_3}$ as indicated in the respective plots. The black points 
indicate points ruled out from direct chargino-neutralino searches. The grey shaded region represents the $95\%$ exclusion region 
around the observed limit.}
\label{fig:excl_mzp_lam01}
\end{figure}
%%%%%%%%%%%%%%%%%%%%%%%%%%%%
It is evident that for small enough $g^{\prime}Q_{E_3}$, even sub-TeV $\mzpr$ is allowed from $\zpr\to\tau\tau$ searches. However, some of this 
parameter space may already be excluded from neutralino/chargino search results at the LHC. Since the bino and wino soft mass parameters
are decoupled from the rest of the spectrum, the LSP can be either a singlino or higgsino. Depending on the nature of the LSP, the 
exclusion limits on the LSP-NLSP masses can be distinctly different. The black points in Fig.~\ref{fig:excl_mzp_lam01} represent 
these excluded regions. The region below the $\zpr$ exclusion limit remains unaffected from the neutralino-chargino searches.  
The region with $\mzpr\lesssim 500$ GeV merits a closer look since some of the neutralino-chargino masses are expected to be light 
enough to be produced in abundance at the LHC. It turns out that all the allowed points shown in the figure have very small LSP-NLSP 
mass gap and hence may avoid detection. We checked some sample points from these regions 
through CheckMATE-2.0.24 \cite{Drees:2013wra,Dercks:2016npn} that they are indeed allowed from latest neutralino-chargino search 
results constraints \cite{Aad:2014vma,Aaboud:2017leg,Aaboud:2018sua}.
%%%%%%%%%%%%%%%%%%%%%%%%%%%%
%%%%%%%%%%%%%%%%%%%%%%%%%%%%
\begin{figure}
%\begin{center}
\subfigure[]{
\includegraphics[width=5.2cm,height=5.5cm]{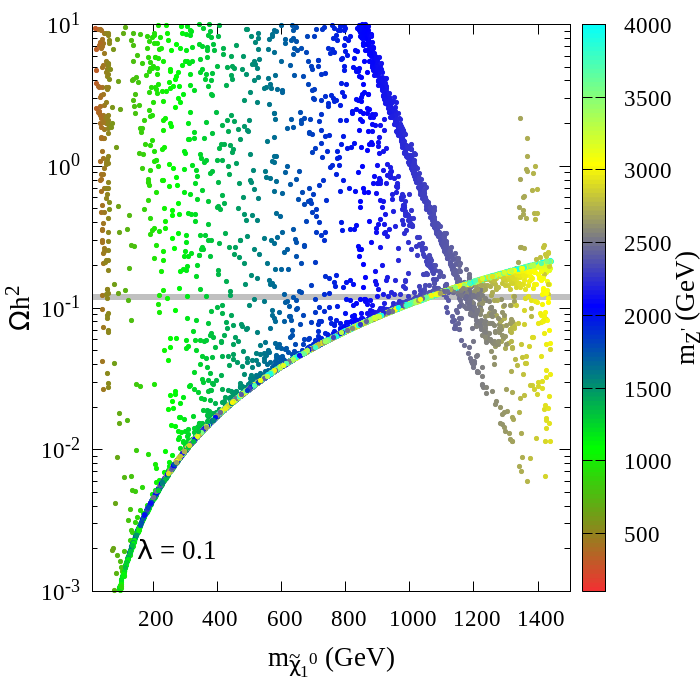}}
\subfigure[]{
\includegraphics[width=5.2cm,height=5.5cm]{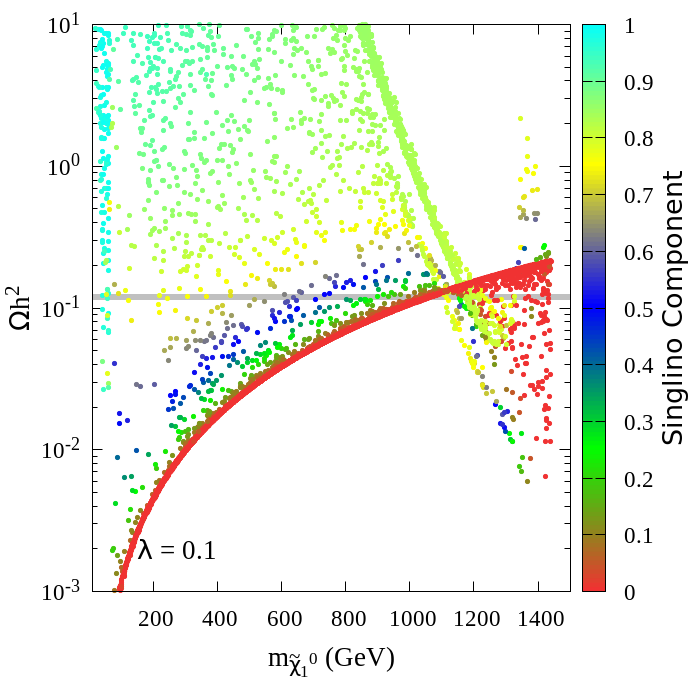}}
\subfigure[]{
\includegraphics[width=5.2cm,height=5.5cm]{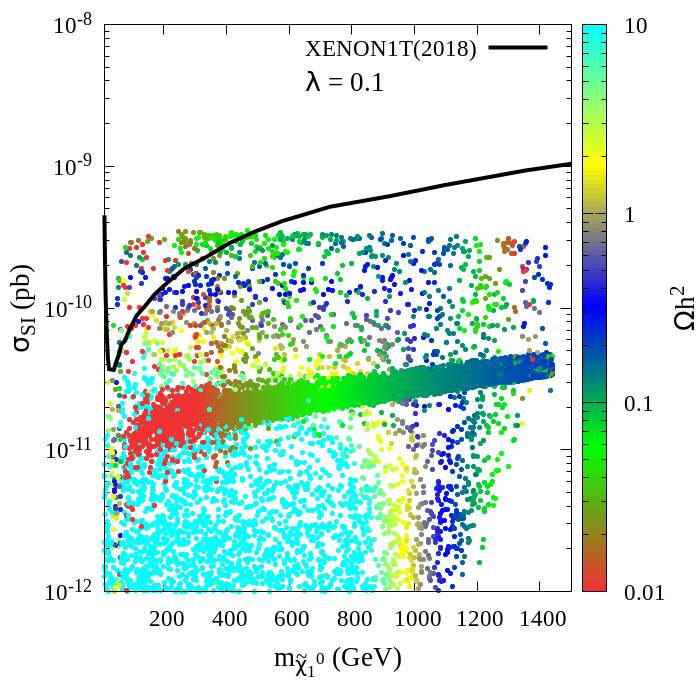}}
\caption{Distribution of the relic density as a function of the LSP mass with the color gradient representing the variation of  
$\mzpr$ (a) and the abundance of the singlino component in the LSP neutralino (b). Panel (c) shows the distribution of 
direct detection cross-section as a function of LSP mass with the color gradient representing relic density. 
All the distributions are for $\lambda=0.1$.} 
\label{fig:relic_mch_lam01}
\end{figure}
%%%%%%%%%%%%%%%%%%%%%%%%%%%%

The distribution of relic density and direct detection cross-section of the LSP in this scenario are shown in Fig.~\ref{fig:relic_mch_lam01}. 
The $\lambda$ parameter being smaller, 
one would expect the effective $\mu$ term to be smaller in comparison with the previous case. Hence there is 
a large region of parameter space where the LSP is purely higgsino-like or a well-mixed singlino-higgsino state. The abundance of 
the red points in Fig.~\ref{fig:relic_mch_lam01}(b) illustrates this feature. As expected, for sub-TeV neutralino states, these points result 
in underabundance of relic density due to too much co-annihilation. However, there is also a significant amount of parameter space 
where the points produce just the correct relic abundance with well-mixed singlino-higgsino LSP states, as represented by the blue 
and green points. These points are also safe from $\zpr$ searches with $\mzpr\gtrsim 1.5$ TeV as can be 
observed from Fig.~\ref{fig:relic_mch_lam01}(a). The direct-detection constraint is not too severe in this case. For the pure 
higgsino LSP (indicated by the red points in Fig.~\ref{fig:relic_mch_lam01}(b), the contributions from two higgsino components 
cancel each other. Singlino-higgsino admixture produces larger $\sigma_{SI}$, but beyond $\mlsp > 500$ GeV, the parameter space is safe 
from the XENON limit. 

In the next section we present some representative benchmark points with the input parameters and resulting mass spectra and 
decay branching ratios. For large $\lambda$ we observed that only the singlino LSP state lies below $\mzpr$ and therefore, 
the SUSY decay mode of the $\zpr$ is completely invisible. For smaller $\lambda$, as $g^{\prime}$ increases, there is more mixing between the singlino 
and higgsino states and as a result additional neutralino-chargino states start to appear in between the $\zpr$ and the LSP. Now 
$\zpr$ may decay into $\widetilde\chi_i^0\widetilde\chi_j^0$ or $\widetilde\chi_1^{\pm}\widetilde\chi_1^{\mp}$ states that eventually 
yield dilepton or trilepton final states. In principle, a four lepton final state is also possible when a pair of heavier 
neutralinos produced from $\zpr$ decay via the $\ell\bar\ell\widetilde\chi_1^0$ mode.
%%%%%%%%%%%%%%%%%%%%%%%%%%%%
%%%%%%%%%%%%%%%%%%%%%%%%%%%%%%%
\subsection{Benchmark Points}
\label{subsec:bm}
%%%%%%%%%%%%%%%%%%%%%%%%%%%%%%%
From the discussion above, the relevant parameter region can be represented by three different classes of benchmark points.
\begin{itemize}
\item {\bf Class-I:} The masses are aligned in such a way that the $\zpr$ can decay into both the higgsino and singlino type 
 neutralino-chargino states. Thus there are three neutralinos and one chargino lying below $\mzpr$ and there is a sizeable mass 
 gap between LSP singlino and NLSP higgsino states, such that the resulting decay leptons can be hard enough. This class of points is 
 shown in Fig.~\ref{fig:excl_mzp_lam01}.
\item {\bf Class-II:} The hierarchy of masses are similar as in Class-I, except for the fact that the LSP can be either 
 singlino or higgsino dominated or a well-mixed state. The NLSP-LSP mass gap is small 
 and thus the final state leptons are softer. This class of points is also shown in Fig.~\ref{fig:excl_mzp_lam01}.
\item {\bf Class-III:} Only the LSP state is lighter than the $\zpr$. The LSP can either be a singlino or higgsino. The NLSP 
 has a mass that kinematically forbids $\zpr$ to decay into any chargino or neutralino pairs. Otherwise, it is simply heavier than $\zpr$. In this case, the 
 $\zpr$ has a large invisible branching ratio. This class of points is shown in Fig.~\ref{fig:excl_mzp_lam06}.
\end{itemize}

In the next section, we shall concentrate only on benchmark points belonging to Class-I and Class-II since the $\zpr$ in BP5 has no visible 
decay into SUSY particles.
%%%%%%%%%%%%%%%%%%%%%%%%%%%%
\begin{table}[h!]
\begin{center}
\begin{tabular}{||c||c|c||c|c||c||} 
\hline
\multicolumn{1}{||c||}{Parameters} &
\multicolumn{2}{|c||}{Class-I} & 
\multicolumn{2}{|c||}{Class-II} &
\multicolumn{1}{|c||}{Class-III}\\
\cline{2-6}
 \& BR    & BP1 & BP2 & BP3 & BP4 & BP5  \\
\hline\hline
${\rm tan}\beta$   &  10.0    & 11.6  &14.36   &10.12    &7.25    \\
$Q_{E_3}$          &  0.5     & 0.5  &0.65    &0.65     &0.91    \\
$g^{\prime}$       &  0.3     & 0.3  &0.3     &0.3      &0.3     \\
$\lambda$          &  0.1     & 0.1  &0.1     &0.1      &0.6     \\
$v_S$  (GeV)       &  9203.0  & 10562.0  &8590.3  &8840.0   &8745.9  \\
\hline
$m_{h_1}$  (GeV)                    & 124.8      & 124.7  &125.7   &125.0   &126.2     \\
$m_{h_2}$  (GeV)                    & 1381.1     & 1584.9  &1670.0  &1734.8  &2741.2   \\
$\mzpr$    (GeV)                    & 1379.7     & 1572.8  &1670.6  &1735.6  &2400.2   \\
$m_{\widetilde\chi_1^0}$  (GeV)     &  428.5     & 543.1  &600.1   &633.0   &1075.0    \\
$m_{\widetilde\chi_2^0}$   (GeV)    &  666.3     & 764.3  &622.9   &640.1   &3713.4    \\
$m_{\widetilde\chi_3^0}$   (GeV)    &  668.7     & 766.7  &630.0   &656.4   &3732.9    \\
$m_{\widetilde\chi_1^{\pm}}$ (GeV)  &  667.4     & 765.4  &624.0   &641.1   &3717.9    \\
\hline
BR($\zpr\to\tau\bar\tau$)                                   &  0.45     & 0.48  &0.35  &0.35   &0.65    \\
BR($\zpr\to\widetilde\chi_1^0\widetilde\chi_1^0$)           &  0.18     & 0.14  &0.05  &--     &0.06    \\
BR($\zpr\to\widetilde\chi_i^0\widetilde\chi_j^0$)           &  0.09     & 0.09  &0.24  & 0.29  &--      \\
BR($\zpr\to\widetilde\chi_1^{\pm}\widetilde\chi_1^{\mp}$)   &  0.09     & 0.08  &0.20  &0.20   &--      \\
BR($\zpr\to q_k\bar q_k$)                                   &  0.19     & 0.21  &0.16  &0.16   &0.29    \\
%\hline
%$\sigma(pp\to\zpr)$ (fb)      &  22.07     &   &15.28  &14.19   &   & \\
\hline\hline
\end{tabular}
\caption{Relevant masses and branching ratios of the benchmark points studied here. Here $i(j)\equiv 2,3$ and $k\equiv 1,2,3$. }
\label{tab:bp_anls}
\end{center}
\end{table} 
%%%%%%%%%%%%%%%%%%%%%%%%%%%%
\section{Collider Analysis}
\label{sec:coll}
%%%%%%%%%%%%%%%%%%%%%%%%%%%%
So far, we observed that for small enough values of $Q_{E_3}$, the $\zpr$ can easily avoid detection in the conventional search channels at 
the LHC. Under such circumstances, although the $\zpr$ has a significantly large decay branching ratio into the $\tau\tau$ mode, 
the production cross-section is simply not large enough for $\zpr$ to be detected. Within a SUSY framework, however, the $\zpr$ has 
additional decay modes which can be explored. Lowering the $Q_{E_3}$, $g^{\prime}$ and $v_S$ parameters results in small $\zpr$ 
masses. At the same time these also lower the singlino mass. Additionally, for small $\lambda$ choices, there can be higgsino-like 
neutralino and chargino states lying below the $\mzpr$. Hence the $\zpr$ can easily decay into 
$\widetilde\chi_i^{\pm}\widetilde\chi_j^{\mp}$ and $\widetilde\chi_i^0\widetilde\chi_j^0$ modes. 
Note that in principle, the bino and wino dominated states can also easily have masses lying in between $\zpr$ and the LSP. 
This can result in a rich cascade decay starting from the resonance production of $\zpr$, but the constraints on bino and wino-like 
neutralino-chargino states are comparatively more severe\footnote{In that case, quite a large portion of the parameter region with sub-TeV $\mzpr$ in that case will be 
discarded based on the bino-wino search results. Thus it is safe to assume that the bino and wino mass parameters are much heavier than 
$\mzpr$} \cite{ATLAS:2017uun,Aaboud:2017leg,Sirunyan:2018iwl}. The higgsino LSP scenario is understandably the least constrained one since its production 
cross-section is comparatively smaller and the NLSP-LSP states are mass degenerate. Note however that the constraints on binos and winos are 
not that robust, so looking at light binos and winos in this model might prove an interesting avenue to pursue in future work.

Depending on the number of neutralino-chargino states lying below $\mzpr$, the observable final states can be quite different. A large 
parameter space discussed so far has either the singlino or the higgsino-dominated states accessible to the $\zpr$ decays. In that case, 
the $\zpr$ decays invisibly into these channels and the $\tau\tau$ decay mode is the one more likely to be seen first. If both the singlino 
and the higgsino states lie below $\mzpr$, from the cascade decay one can expect to obtain two or more leptons in the final state associated with missing 
energy.  Therefore, we use the multilepton search results from the LHC to ascertain the sensitivity of this search strategy for 
probing $\zpr$ in the present scenario. We then proceed to make an estimate of the LHC sensitivity at high luminosity. Note that the sensitivity 
of these multilepton search strategies in probing the present scenario is likely to vary depending on the mass difference between the light 
neutralino-chargino states \cite{ATLAS:2017uun,Aaboud:2017leg,Sirunyan:2018iwl}. Two sets of kinematic cuts are therefore chosen in such a way so as to gain 
maximum possible sensitivity for the different sets of benchmark points.

For our collider analysis, we have used MadGraph5 \cite{Alwall:2011uj,Alwall:2014hca} to generate events at the parton level which are 
subsequently passed through PYTHIA8 \cite{Sjostrand:2006za,Sjostrand:2014zea} for decay, showering and hadronisation. nn23lo1 parton distribution function 
\cite{Ball:2012cx,Ball:2014uwa} has been used while simulating signal as well as SM background events. MLM matching \cite{Hoche:2006ph,Mangano:2006rw} 
scheme has been used for production channles with light jets at the parton level.  We have used anti-kt algorithm \cite{Cacciari:2008gp} 
in FastJet \cite{Cacciari:2011ma} for construction of jets and Delphes \cite{deFavereau:2013fsa,Selvaggi:2014mya,Mertens:2015kba} for 
detector simulation. Finally, we perform our analysis in CheckMATE \cite{Drees:2013wra,Dercks:2016npn}.  
%%%%%%%%%%%%%%%%%%%%%%%%%%%%
\subsubsection{Cuts for benchmark points class I}
\label{subsubsec:cuts_class1}
%%%%%%%%%%%%%%%%%%%%%%%%%%%%
For this class of benchmark points, apart from di-taus, di-leptons associated with missing transverse energy can be a possible signal. 
Note that contribution to this new signal region for $\zpr$ can only arise from the small branching ratio of its decay into the charginos or neutralino states. 
As can be observed from Table~\ref{tab:bp_anls}, a $\zpr$ branching ratio of 18\% for the decay  is relevant to this case, which is further diminished by the leptonic branching ratio of the
decay  of the gauge bosons. Hence the resultant event rate is expected to be small and the di-tau signal regions is expected to be observed 
first if such a $\zpr$ exists. However, the di-leptonic signal region, if observed further at high luminosity, can serve as a robust hint of 
existence of SUSY. 

The dominant SM background channels for this signal region are $t\bar t +~{\rm jets}$, $t\bar t + V$ ($V = W^{\pm}, Z$), $t\bar t + h$, $VV$, 
$VVV$ and $Z +~{\rm jets}$. We set the following criteria for selection of the final state.
\begin{itemize}
 \item {\bf C1:} The final state must have two opposite-sign different flavor leptons. The transverse momentum, $p_T$ of the leading and 
 sub-leading leptons are required to be more than 25 GeV and 20 GeV respectively. 
 \item {\bf C2:} No central light jets with $p_T > 40$ GeV and $|\eta| < 2.4$. 
 \item {\bf C3:} No central $b$-tagged jets with $p_T > 20$ GeV and $|\eta| < 2.4$. 
 \item {\bf C4:} The invariant mass of opposite-sign di-leptons pair, $m_{\ell\ell}$ has to be away from the Z-boson mass ($m_Z$), i.e. 
 $|m_{\ell\ell} - m_Z| > 10$ GeV. 
 \item {\bf C5:} The missing transverse energy, $\slashed{E}_T$ has to be more than 200 GeV.
 \item {\bf C6:} The stransverse mass, $m_{T_2} = {\rm min}_{\vec{q}_T}[{\rm max}(m_T(\vec{p}_T^{\ell_1}, \vec{q}_T), m_T(\vec{p}_T^{\ell_2}, \vec{p}_T^{\rm miss} - \vec{q}_T))]$,  
 should be more than 150 GeV. Here $m_T$ is given by $m_T (\vec{p}_T, \vec{q}_T) = \sqrt{2(p_T q_T - \vec{p}_T . \vec{q}_T)}$. 
\end{itemize}
%%%%%%%%%%%%%%%%%%%%%%%%%%%%
\subsubsection{Results for benchmark points class I}
\label{sec:res_class1}
%%%%%%%%%%%%%%%%%%%%%%%%%%%%
\begin{table}[h!]
\begin{center}
\begin{tabular}{||c||c|c|c|c|c|c||}
\hline
\multicolumn{1}{||c||}{Channels} &
\multicolumn{6}{|c||}{Cross-section (fb)} \\
\cline{2-7}
& {\bf C1} & {\bf C2} & {\bf C3} & {\bf C4} & {\bf C5} & {\bf C6} \\
\hline\hline
{\bf BP1} & 1.008& 0.572& 0.544& 0.504& 0.207& 0.007 \\
{\bf BP2} & 0.593& 0.330& 0.313& 0.291& 0.133& 0.005\\
\hline
$t\bar t +~{\rm jets}$ & 13823.5& 7756.9& 423.1& 406.6& 6.535&-- \\
$t\bar t + X$ & 85.992& 37.568& 0.546& 0.497& 0.032& --  \\
$VV$ & 1755.233& 1362.872& 1343.398& 1086.805& 1.104& 0.003 \\
$VVV$ & 15.021& 4.119& 2.966& 2.430& 0.117& 0.012 \\
%$Z +~{\rm jets}$ & $1.53\times 10^7$& $1.36\times 10^7$& $1.35\times 10^7$& $6.41\times 10^5$& --&-- \\
\hline\hline
\end{tabular}
\caption{Cutflow table for signal and SM background channels for BP1 and BP2 benchmarks.}
\label{tab:res_SRs}
\end{center}
\end{table}
%%%%%%%%%%%%%%%%%%%%%%%%%%%%
In this case the gauge boson production channels are the most dominating contributors to the background. Cuts {\bf C5} and {\bf C6} 
effectively reduce these contributions. Cuts {\bf C2} and {\bf C3} are particularly helpful in reducing the backgrounds from top 
production channels which are further reduced by {\bf C6}. The requirement that the leptons need to be different flavor is helpful 
in reducing the leptons arising from the $Z$ boson decay. Including the same-flavor lepton pairs enhances the signal rate, but the 
background contribution specially from $VV$ production channel becomes too large even in the presence of cut {\bf C4}. The large $m_{T_2}$ 
cut proves to be most effective in getting rid of the background although it also reduces the signal events to a large extent. Overall, 
one requires an integrated luminosity of $\sim 1.4~{\rm ab}^{-1}$ and $\sim 2.6~{\rm ab}^{-1}$ to exclude (or to achieve $2\sigma$ 
statistical significance) {\bf BP1} and {\bf BP2} 
respectively\footnote{To compute statistical significance we have used ${\mathcal S}=\sqrt{2(S+B){\rm Log}(1+\frac{S}{B})-S}$.}. 
To achieve a $3\sigma$ statistical significance one requires $\sim 3.1~{\rm ab}^{-1}$ and $\sim 6~{\rm ab}^{-1}$ integrated 
luminosity respectively. The high-luminosity LHC is expected to reach an integrated luminosity of 3 ab$^{-1}$. 
There is also one high-energy LHC proposal that will operate at 27 TeV and is expected to reach 15 ab$^{-1}$ luminosity.
%%%%%%%%%%%%%%%%%%%%%%%%%%%%
\subsubsection{Cuts for benchmark points class II}
\label{sec:cuts_class1}
%%%%%%%%%%%%%%%%%%%%%%%%%%%%
Benchmark points under class II have smaller NLSP-LSP mass gap and as a result we cannot use a hard $m_{T_2}$ cut to reduce background 
contributions effectively. Instead, we devised the cuts in a way so that the softness of the leptons and the large missing energy can be 
utilised to reduce the SM events. The criteria used here are:
\begin{itemize}
 \item {\bf D1:} The final state must have two opposite-sign leptons with their $p_T$ within the range $[5, 30]$ GeV. For electrons,
 $|\eta_e| < 2.4$ and for muons, $|\eta_{\mu}| < 2.5$. 
 \item {\bf D2:} At least one light jet with $p_T > 25$ GeV and $|\eta| < 2.4$. 
 \item {\bf D3:} No central $b$-tagged jets with $p_T > 25$ GeV and $|\eta| < 2.4$.
 \item {\bf D4:} Missing energy, $\slashed{E}_T > 250$ GeV. 
 \item {\bf D5:} Transverse mass, $m_T(\ell_i, \slashed{E}_T) < 70$ GeV, where $i = 1, 2$.
 \item {\bf D6:} Invariant mass of opposite-sign lepton pair, $4 < m_{\ell\ell} < 25$ GeV.
\end{itemize}
%%%%%%%%%%%%%%%%%%%%%%%%%%%%
%%%%%%%%%%%%%%%%%%%%%%%%%%%%
\begin{table}[h!]
\begin{center}
\begin{tabular}{||c||c|c|c|c|c|c||}
\hline
\multicolumn{1}{||c||}{Channels} &
\multicolumn{6}{|c||}{Cross-section (fb)} \\
\cline{2-7}
& {\bf D1} & {\bf D2} & {\bf D3} & {\bf D4} & {\bf D5} & {\bf D6} \\
\hline\hline
{\bf BP3} & 0.168& 0.083& 0.072& 0.010& 0.003& 0.003\\
{\bf BP4} & 0.025& 0.013& 0.011& 0.004& 0.002& 0.002\\
\hline
$t\bar t +~{\rm jets}$  & 2749.1& 2670.6& 709.87& 4.392& 0.088& 0.003 \\
$t\bar t + X$ & 11.56& 11.48& 2.208& 0.047& 0.002& -- \\
$VV$ & 339.51& 73.52& 67.14& 0.753& 0.305& 0.005 \\
$VVV$ & 1.193& 0.937& 0.737& 0.017& 0.006& 0.001 \\
%$Z +~{\rm jets}$ & & & & & &-- \\
\hline\hline
\end{tabular}
\caption{Cutflow table for signal and SM background channels for BP3 and BP4 benchmarks.}
\label{tab:res_SRs}
\end{center}
\end{table}
%%%%%%%%%%%%%%%%%%%%%%%%%%%%
In order to reduce the background contributions from the gauge boson production channels, we put strict restrictions on the transverse mass 
of the charged leptons and missing energy. This cut, combined with the large missing energy one, effectively reduce the background contributions. 
A further restriction on the invariant mass of the same-flavor lepton pairs ensures that even such a small signal rate can be observed at the high 
luminosity LHC. Reducing the $VV$ background proves to be difficult in this case. Demanding the presence of at least one hard jet coupled with a large 
missing energy cut is useful to this effect. Moreover, demanding a small invariant mass window ({\bf D6}) 
reduces this background effectively. The resulting statistical significance of this class of benchmark points is understandably small due to smaller 
production cross-section of the signal. {\bf BP3} and {\bf BP4} require an integrated luminosity of $\sim 4~{\rm ab}^{-1}$ and $\sim 10~{\rm ab}^{-1}$ 
respectively to achieve a $2\sigma$ statistical significance. To obtain $3\sigma$, one requires $\sim 10~{\rm ab}^{-1}$ and $\sim 22~{\rm ab}^{-1}$ 
respectively. 

Note that the large luminosity requirement for {\bf BP4} observation makes it most unlikely to be probed at the LHC in the above-mentioned signal region mainly 
because of the very small NLSP-LSP mass gap ($\sim 7~{\rm GeV}$). For these kind of points, one can consider probing a mono-jet signal region where 
one of the initial-state-radiation (ISR) jets is tagged \cite{Aaboud:2017leg}. However, this signal region has large a hadronic background that is almost 
impossible to get rid of against such a small signal rate. 
%%%%%%%%%%%%%%%%%%%%%%%%%%%%
\section{Conclusion}
\label{sec:concl}
%%%%%%%%%%%%%%%%%%%%%%%%%%%%
We have considered a scenario where the MSSM is extended by one additional $U(1)^{\prime}$ gauge group. The $U(1)^{\prime}$ charges for the fermions and Higgs bosons are family dependent, 
which allows for  cancellation of anomalies without the introduction of exotic states, and leads to interesting phenomenological consequences. We consider one possible 
solution to all the anomaly cancellation conditions in such a way that all the $U(1)^{\prime}$ charges can be written in terms of $Q_{E_3}$,  the 
corresponding $U(1)^\prime$ charge for $E_3^c$. The resulting charge assignments require one to introduce non-holomorphic 
SUSY breaking Lagrangian to the theory in order to avoid massless fermions. They also forbid the $\zpr$  decay into an electron or muon pair at the tree level, which circumvents the most stringent constraint on $\mzpr$. In absence of these decay modes the restriction on $\mzpr$ 
arises from $Z^\prime$ decay into $\tau\bar\tau$ final state, which is understandably much weaker. The signal cross-section is also dependent on the choice of $U(1)^{\prime}$ charges 
and other possible decay modes of $\zpr$. In the framework of SUSY, there can be some other decay modes. Here we have explored the possibility of its decay into 
multiple chargino and neutralino states that can give rise to observable leptonic signals at high luminosity LHC. Since we are working within a $R$-parity conserving 
framework, the LSP neutralino can be a DM candidate. A non-standard candidate for LSP such as a singlino or a higgsino arises naturally in this framework if one considers a light $\zpr$. 
Hence we restricted ourselves to these two possibilities and performed a scan of the parameter space by varying $\lambda$,   $\tan \beta$, and $g^\prime Q^\prime$, where  $Q^\prime \equiv Q^\prime_{E_3}$ 
to highlight the available parameter space taking into account both 
the collider and DM constraints. We proceed to study two possible signal regions with a pair of opposite-sign leptons in the final state with different set of kinematic 
cuts chosen suitably depending on the varying NLSP-LSP mass gap. We observed that even in the presence of these 
additional decay modes, the di-tau final state is likely to be observed first and if it so happens, one can use the the leptonic signal regions as confirmatory channels. 
In the present framework, any observation of such leptonic signals at high luminosity will also indicate the presence of SUSY.  
%%%%%%%%%%%%%%%%%%%%%%%%%%%%
\section*{Acknowledgements}
%%%%%%%%%%%%%%%%%%%%%%%%%%%%
MF acknowledges NSERC for partial financial support under grant number SAP105354. SM and KH acknowledge H2020-MSCA-RICE-2014 grant no. 645722 (NonMinimal Higgs). 
%%%%%%%%%%%%%%%%%%%%%%%%%%%%%%%%
\bibliography{nonuniZp}{}
%%%%%%%%%%%%%%%%%%%%%%%%%%%%%%
\end{document}